# Global interpretation of LHC indications within the Georgi-Machacek Higgs model

François Richard[1]

Université Paris-Saclay, CNRS/IN2P3, IJCLab[2], 91405 Orsay, France
_________________________________________________________________________

*Abstract: Following various LHC indications for new scalars, an interpretation of these is given in terms of the Georgi-Machacek (GM) model. On top of the confirmed SM Higgs boson, there are indications for a light Higgs at 96 GeV, for a CP-odd boson at 400 GeV, A(400), and for a heavy Higgs boson at 660 GeV. An extension of the GM is needed to interpret the fermion couplings of A(400). Potentially interesting deviations are also observed in the ttW cross-section measurement, which naturally fit into this picture. None of them crosses the fatidic five s.d. evidence but the addition of these effects, consistent with GM, suggest that there are good hopes for solid discoveries at HL-LHC, which should boost the motivation for future machines. The GM model also provides a useful framework to estimate the rates expected for various channels at an e+e- collider, together with the range of energies needed. ILC performances are used for a quantitative estimate of these rates for the prominent channels.*

## Introduction

A collection of indications for BSM physics from ATLAS and CMS was described in two previous notes [1, 2]. The present note provides an updated version of this analysis and an attempt to find a **consistent theoretical interpretation** of these effects.

It is well understood that the SM cannot be the last word since, among many examples, it is unable to provide the necessary inputs to understand our world, in particular to understand baryogenesis.

_____________________________

[1] richard@ijclab.in2p3.fr
[2] Laboratoire de Physique des 2 Infinis Irène Joliot-Curie

(N)MSSM with CPV, could remedy to this situation but it seems that, in spite of its apparent triumph in predicting the Higgs mass, it has lost its basic motivation, which is SUSY, insuring an appropriate cancellation of the quartic mass divergences for **elementary scalars**. A possible way out is to move to **compositeness** which also provides motivations for light scalars, as PNGB (Pseudo Nambu-Goldstone Bosons) particles of an unknown broken symmetry with no precise predictions for the mass spectrum of the new particles. Composite models also require structuring the new scalars into weak isospin multiplets, similar to what does MSSM within SUSY. This allows passing the precision test for the S,T,U variables.

Both the MSSM framework and its composite extensions predict isodoublets and isosinglets and we have seen [1,2] that they enter in conflict with some of the indications provided by LHC results.

At variance with these ideas, Georgi and Machacek (GM) [3] have offered a radically different group structure for the Higgs isomultiplets, quote: "*The possibility that representations containing double charged scalars may participate in the spontaneous breakdown of the SU(2)xU(1) symmetry of electroweak interactions*".

The usual dogma, usually assumed for the Higgs sector, states that to satisfy the identity **$M_Z \cos\theta_W / M_W = 1$** (up to loop corrections), one can only allow for isosinglets and isodoublets. The GM model allows for higher isospin states, in such a way as keeping above identity at the tree level, hence the possibility of accommodating exotic scalars with double charge. This result can be achieved without fine-tuning by implementing the custodial symmetry in the Higgs potential [4].

It turns out, as will be shown in this note, that this model offers viable solutions for most inconsistencies between LHC findings and the orthodox extensions of the SM. Examples of this are:

- Observation of a heavy scalar in the ZZ mode, while MSSM predicts a decoupling
- Predominant production of H(660) through VBF fusion, ZZ/WW->H(660), in the absence of coupling to heavy fermions which forbids gluon-gluon fusion ggF
- Indication for a CP-odd scalar, with A(400)->Zh, while MSSM predict a decoupling of this mode
- Indication for a 50% excess in ttW, which naturally emerges within this model through the VBF process WZ->H+(660)->A(400)W+ with A(400)->tt
- A smaller excess in ttZ which is understood as due to the smaller production rate for ZZ->H(660)->A(400)Z  and a larger SM cross-section

One may object to this model an absence of doubly charged signals into the W+W+ mode, but, as will become clear, this can be interpreted as a dominance of complex decay modes like H+(400)W+, much harder to detect at LHC.

Could h(96), indicated by LEP2 and CMS, belong to this GM structure? This seems a priori plausible since the model predicts a second isosinglet together with h(125), but we will see that in such a case one should observe the transition A(400)->h(96)Z with a rate incompatible with LHC findings. I will therefore assume that the second isosinglet is heavier, presumably heavy enough that the decay into h(125)h(125) becomes dominant.

Section II, after recalling these **LHC anomalies**, gives their interpretation within the GM model, with more details in the Appendix. Recently, ATLAS has updated its search for H->ZZ and observed an indication in the VBF channel at the relevant mass. The absence of any indication into WW favours the GM model that predicts WW/ZZ~0.5 for a 5-plet scalar.

Section III is about **cascade decays predicted** by the GM model and possible signals at LHC. A 50% excess in the ttW channel, observed by ATLAS and CMS, is interpreted within the same model. Possibilities offered by lepton tagging are also discussed and illustrated by an ATLAS search. Various anomalies observed in **multileptonic states**, in some cases accompanied by b-jets, are recalled which may, at the qualitative level, lead to a similar interpretation. An **alternate explanation** of these states is offered by a model where two light scalars cascade into each other, the lightest being indicated by a recent analysis of LHC data.

Section IV will discuss possible **extensions of the GM model** needed to interpret some apparent contradictions found for the coupling of A(400) to top and tau pairs (cf. III.1 of the Appendix). The flexibility offered by the so-called **Aligned-Two-Higgs-Doublet-Scheme** are used to interpret the fermionic couplings of A(400).

Section V examines what can be hoped for at **HL-LHC** within the GM model.

Section VI does the same for **future e+e- colliders**.

## II.   A Georgi-Machacek interpretation of LHC anomalies

### II.1 LHC anomalies

Let me briefly recall what are these anomalies:

- Indications for a scalar into two photons at 96 GeV from CMS
- Several indications (top pairs, tau pairs, Zh(125)[3]) for a pseudo scalar at 400 GeV . When combined statistically, these indications amount to ~6 s.d.
- A ~4.3 s.d. local excess at ~660 GeV in the golden mode ZZ into four leptons, obtained by combining CMS and ATLAS data
- This signal corresponds to a HZZ coupling incompatible with MSSM as explained below

Recall that it is not trivial to interpret the various indications for a resonance at 400 GeV:

- gg->A(400)->tt interferes strongly with the QCD background, which renders the extraction of the cross-section delicate
- For A(400), a top Yukawa coupling close to 1 can explain the 3.5 s.d. effect seen in CMS, implying, within MSSM, a negligible coupling to b quarks and taus, which does not seem to be the case since one has evidence for associated production A+bjet in the Zh and $\tau\tau$ channels
- A->hZ is incompatible with MSSM which predicts a decoupling of this mode

As already stated in my previous notes, these observations do not fit into the usual MSSM scheme. This was also emphasized in [5] for A->hZ. So far, I had refrained from giving an alternate explanation for these observations, interpreting these inconsistencies as due to the composite origin of these scalars. Here, I will indicate why GM gives a natural framework for some of these observations, even suggesting additional observations, called generically "cascades", which naturally emerge in such a phenomenology.

---

[3] Recently, ATLAS collaboration has updated this search with a negative result on ggF->Zh, while no update on bbA was released, see *ATLAS CONF-2020-043 27 July 2020.*

GM naturally includes A(400) and H5(660). It cannot include h(96) as the second singlet, since it would imply a large A(400)->h(96)Z BR (Branching Ratio), not supported by the data. I will therefore assume that the second singlet, hereafter called h', is heavy. h' cannot explain the ZZ bump at 660 GeV since it decays dominantly into hh and has a BR into ZZ at the % level (see figure 10).

Recently ATLAS [6] has published an update of the 4 leptons search, with the full sample (139 fb-1) analysed. In figure 1, one observes an excess in the mass region of interest.

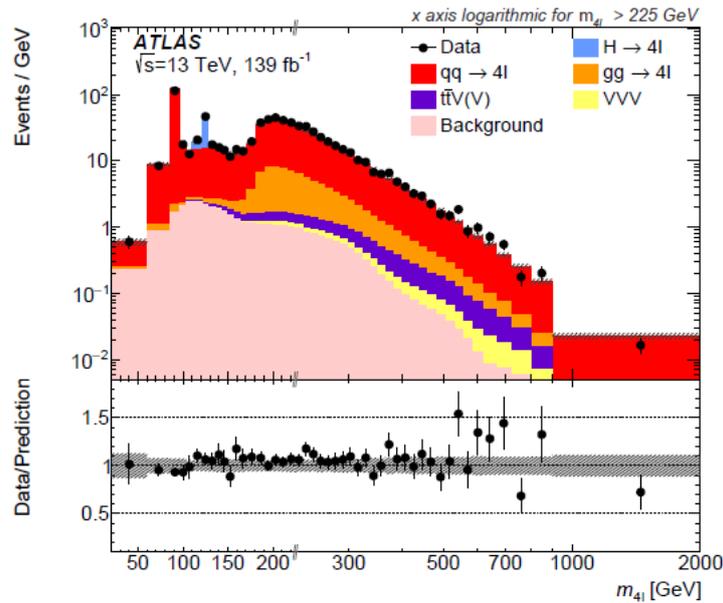

*Figure 1: ATLAS results for the 4 leptons analysis with an integrated luminosity of 139 fb-1 at 13 TeV.*

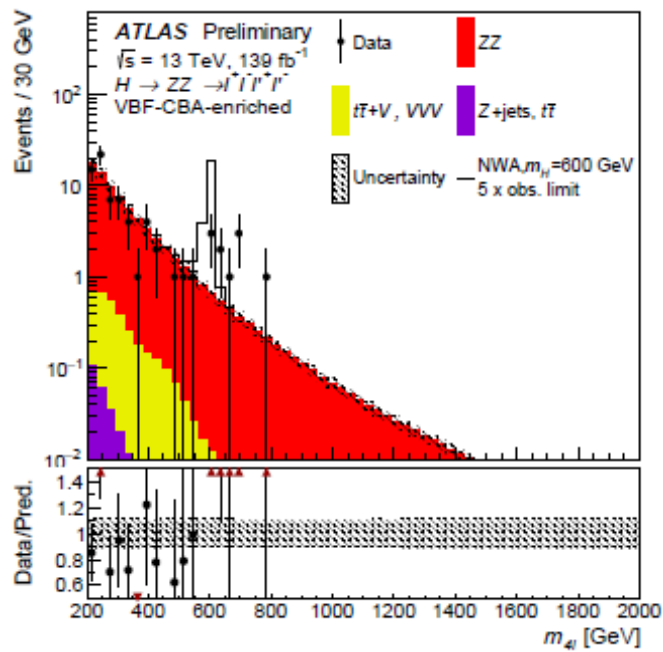

*Figure 2: Cut based analysis of ATLAS for the VBF->ZZ channel into 4 leptons. The predicted signal indicated in the figure is narrower than the expected signals for H5.*

Figure 2 shows the result obtained by ATLAS for the VBF search, with a **cut-based** analysis for the 4-lepton case, suggestive of an excess around 660 GeV. This analysis shows that a VBF selection allows reaching a good signal/background ratio. It was only available in ATLAS-CONF-2020-032 and not reported in the final publication. This measurement allows extracting the partial width $\Gamma_{ZZ}$~30 GeV knowing that the observed total width, see [1], is ~100 GeV.

The large partial width $\Gamma_{ZZ}$~30 GeV, implies that $g^2_{H660ZZ}$~$0.6 g^2_{h125ZZ}$ , which is strictly at variance with sum rule which should be satisfied within MSSM:

$$\sum g^2_{hiZZ} = g^2_{HSMZZ} \text{ knowing that } g^2_{h125ZZ} \sim g^2_{HSMZZ}$$

The signal acceptance, defined as the ratio of the number of reconstructed events after all selection requirements to the total number of simulated events, is found between 30% (15%) and 46% (22%) in the ggF(VBF)-enriched category for the ggF(VBF) production mode depending on the signal mass hypothesis. This means that a large fraction, if not all, of the signal observed in figure 1 could come from VBF as expected in GM.

Note that the two resonances h' and H5(660) could very well overlap and eventually interfere. This effect is however expected to be weak since, as discussed in the Appendix, h'(600) will decay dominantly into h(125)h(125).

## II.2 Indication for a charged Higgs?

In contrast to MSSM, GM predicts a tree-level coupling of H5+ into ZW, which allows **VBF production** of such a particle. The two figures below [7,8] – not considered in my previous notes – provide weak but coincident evidence, at similar masses, for such a mechanism but, contrary to expectation, at a mass clearly below H5(660). ATLAS observes, for the VBF category, a 2.9 standard deviation for mH+=450 GeV. Recall that, within GM, this coupling is only allowed for H5+, which means that there is a mismatch in masses.

The analysed samples, respectively 15.2 fb-1 and 36.1 fb-1, correspond to a small fraction of the presently available luminosity and should be updated.

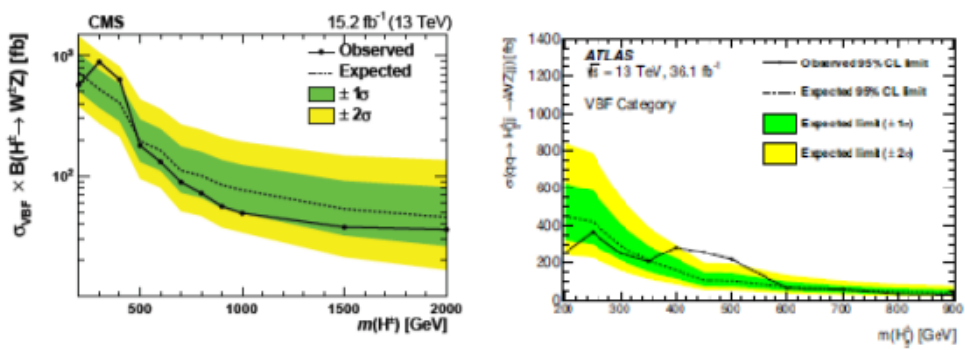

*Figure 3: CMS and ATLAS results on ZW->H+->ZW search.*

## II.3 Quantitative treatment of the GM model

The formalism and the details of the derivation of GM parameters are explained in the **Appendix**. In the present section, I will summarize the line of reasoning and the conclusions.

The GM couplings depend on two mixing angles $\alpha$ and $\theta_H$ and, given the various constraints that need to be fulfilled to describe LHC data, the retained solution is:

**sH=0.50 cH=0.87 sα=-.15  cα=0.99 with the vacuum expectations vχ=43 GeV and vφ=214 GeV.**

This solution satisfies the two mandatory requirements coming from the observed width of H(660) in ZZ interpreted as the neutral H5 of the 5-plet and from the width of A(400) into hZ, interpreted as the neutral H3 of the 3-plet. Moreover the GM model predicts that BR(H5->ZZ)/BR(H5->WW)=2 in contrast to the SM and MSSM models which predict a ratio ½. This implies that the observation of H(660) in WW will be more difficult in such a model.

This solution also follows from existing constraints coming from the SM scalar h(125), which is an isosinglet of the GM solution. Masses from the 3-plet and 5-plet suggested by LHC data also fulfil the S,T parameters as can be seen from figure 4 from [9]. Figure 15 of the Appendix shows that this choice is quite tight. The only missing parameter, as already mentioned, is the mass of the second scalar singlet labelled h'.

The following table summarizes the couplings deduced from this solution and, when relevant, the predicted cross-sections for an e+e- operating at 1 TeV. In section VI, I will give the energy dependence of the rates expected at ILC.

| Type | coupling /SM, MSSM | Numerically | σee fb@1 TeV | e+e- Eth GeV |
|---|---|---|---|---|
| h(125)WW/ZZ | cαcH- 1.63sαsH | 0.98 | 12.0 | 216 |
| h'(600)WW/ZZ | sαcH+1.63cαsH | 0.68 | 1.5 | mh'+mz |
| h(125)tt,bb | cα/cH | 1.14 | | |
| h'tt,bb | sα/cH | 0.17 | | |
| Att,bb,ττ | tanH | 0.58 | | |
| H5WW, H5ZZ | 1.15sH,-2.31sH | 0.57,1.16 | 3 | 751 |
| H5AZ,H5H3+W- | 1.16cH | 1 | 0 | 1060 |
| H5+H3+Z,H5+AW+ | cH | 0.87 | 0 | 1060 |
| Zh(125)A | 1.63(sαcH+0.6cαsH) | 0.28 | 0.4 | 525 |
| Zh'(600)A | 1.63(cαcH- 0.6sαsH) | 1.48 | 0 | mh'+mA |
| h'(600)H3+W- | 1.63(cαcH- 0.6sαsH) | 1.48 | | |
| ZH5+W- | 2sH | 1.0 | 2*2.2 | 740 |
| ZH3+H3- | 1 | 1 | 5.7 | 800 |

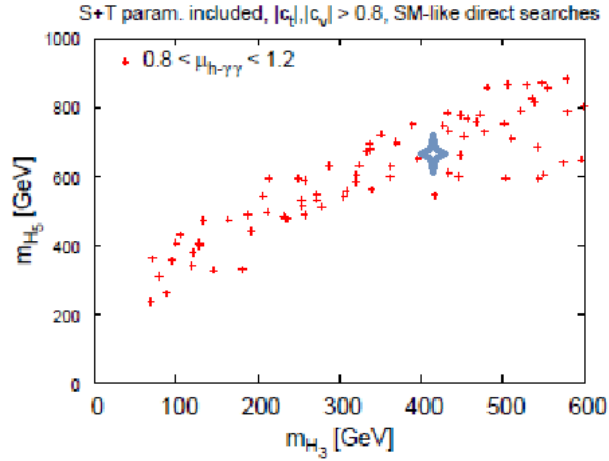

*Figure 4: GM solutions fulfilling S+T and h(125)->2photons constraints. Signal is marked by a cross.*

The following table gives the partial widths for some relevant channels.

The isosinglet h' decays mainly into hh (see Appendix and figure 10).

| Channel | $\Gamma$WW/ZZ GeV | $\Gamma$tt GeV | $\Gamma$Zh(125) GeV | $\Gamma$AZ GeV | $\Gamma$H3W GeV | $\Gamma$tot GeV |
|---------|-------------------|----------------|---------------------|----------------|-----------------|-----------------|
| A(400)  | -                 | 11.1           | 0.38                | -              | -               | 11.5            |
| H5(660) | 15/30             | -              | -                   | 27             | 41              | 110             |

Few remarks are of order:

- The coupling of A(400) to h'Z is ~5 times larger than for h(125)Z which excludes the hypothesis that h(96) could be the missing singlet
- For H5(660), the predicted width comes quite close to the observed width [1,2], ~100 GeV
- Loop contributions (gg, $\gamma\gamma$, Z$\gamma$) are ignored but should not affect the total width result

# III.  Cascades at LHC

In the GM model, one expects that the heavier scalars which, according to our analysis, belong to the 5-plet, will **cascade** into lighter scalars from the triplet and could populate the **topologies ttZ and ttW** studied at LHC. This is also true for the singlet h', but to a lesser degree given the prominent decay h'->2h. The cascade mechanism is pictorially described in the following picture, recalling that particles of the 5-plet are mass degenerate at ~660 GeV. Accordingly, H5 and H5+ will cascade into ZA and WA, which will contribute to ttZ and ttW final states. ttW receives a contribution from H5+ and H5- which are produced by VBF with stronger couplings than the neutral H5.

In the picture below, green is used for particles and processes which are already identified from LHC data.

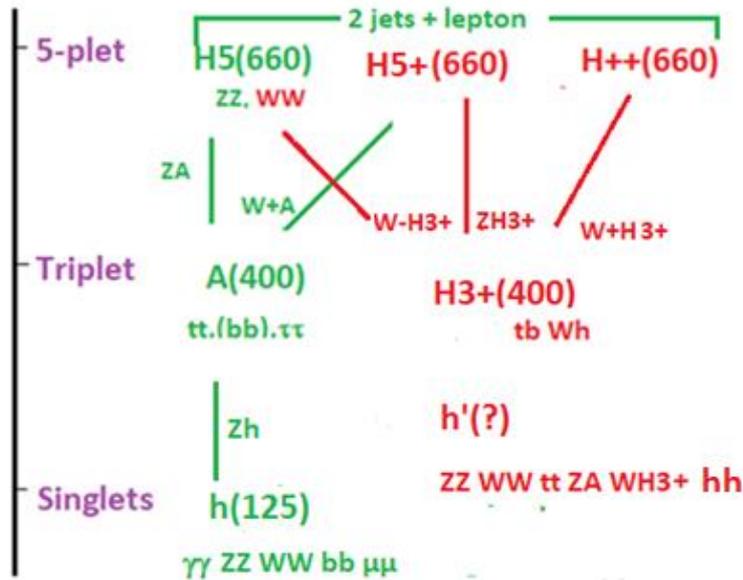

## III.1 ttW and ttZ measurements at LHC

An excess in ttW, was observed, both by ATLAS and CMS, at 8 TeV and confirmed at 13 TeV. Figure 5 shows a 2 s.d. significance for CMS [10].

A similar effect should be observed for ttZ, noting however that the SM cross-section for this process being larger and the expected signal being lower, this effect is harder to observe.

The most recent CMS measurement CMS-PAS-TOP-18-009:

$$\sigma(ttZ)=1.00+0.06-0.05(stat)+0.07-0.06(syst)\ pb$$

agrees with ATLAS-CONF-2020-028:

$$\sigma(ttZ)=1.05+0.05-0.05(stat)+0.09-0.09(syst)\ pb$$

which therefore show the same excess, ~20%, with respect to the NLO+NNLL prediction:

$$\sigma(ttZ)SM=0.863+0.07-0.09(scale)\pm0.03\ pb$$

Given present uncertainties, ttW and ttZ measurements are compatible with this GM interpretation.

An essential point to achieve progress and establish the origin of this mechanism is to realize that the 5-plet is produced by the VBF mechanism, while the SM component comes from ggF for ttZ and from qqF for ttW (hence its lower cross section). **Selecting the VBF topology** should therefore eliminate most of the SM part and confirm this interpretation. To my knowledge, **this measurement has not yet been performed**. Once this selection is operated, one should observe that the top pair masses cluster around 400 GeV. One should also observe a **charge asymmetry** in the selected ttW events produced by ZW fusion which simply reflects the ratio u/d~2 for the valence quarks of the parent protons. Finally, the ttZ topology, with Z into lepton pairs, allows to reconstruct the parent H5 resonance. These statements are true in case one uses fully hadronic top decays since leptonic decays contain unmeasured neutrinos.

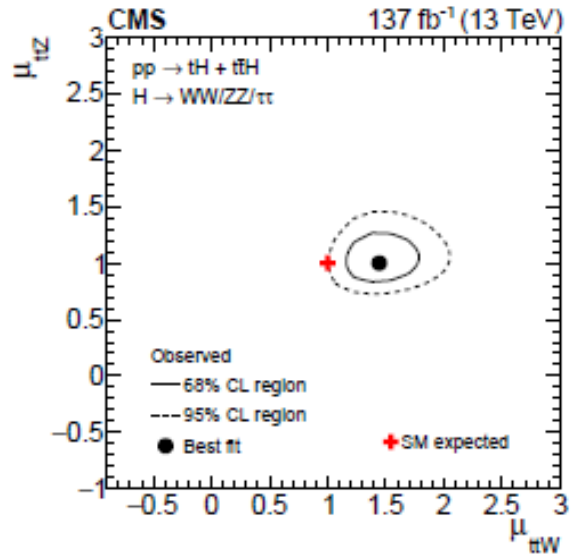

*Figure 5: CMS ttZ and ttW measurements normalized to expectations.*

## III.2 ATLAS search for cascades using lepton tagging

In [11], an attempt has been made to reconstruct resonances decaying hadronically which are accompanied by a lepton with large pT. The cascade mechanisms:

- H5 and h' into AZ and H3+W-
- H5+->AW+,H3+Z
- H5++->H3+W+

can provide such a high pT lepton, which originates from Z/W accompanying an H3 decaying hadronically. The pT selection will be easily fulfilled given that, in the centre of mass of H5, the W/Z energy is E*~210 GeV, which provides a boost to Z/W particles.

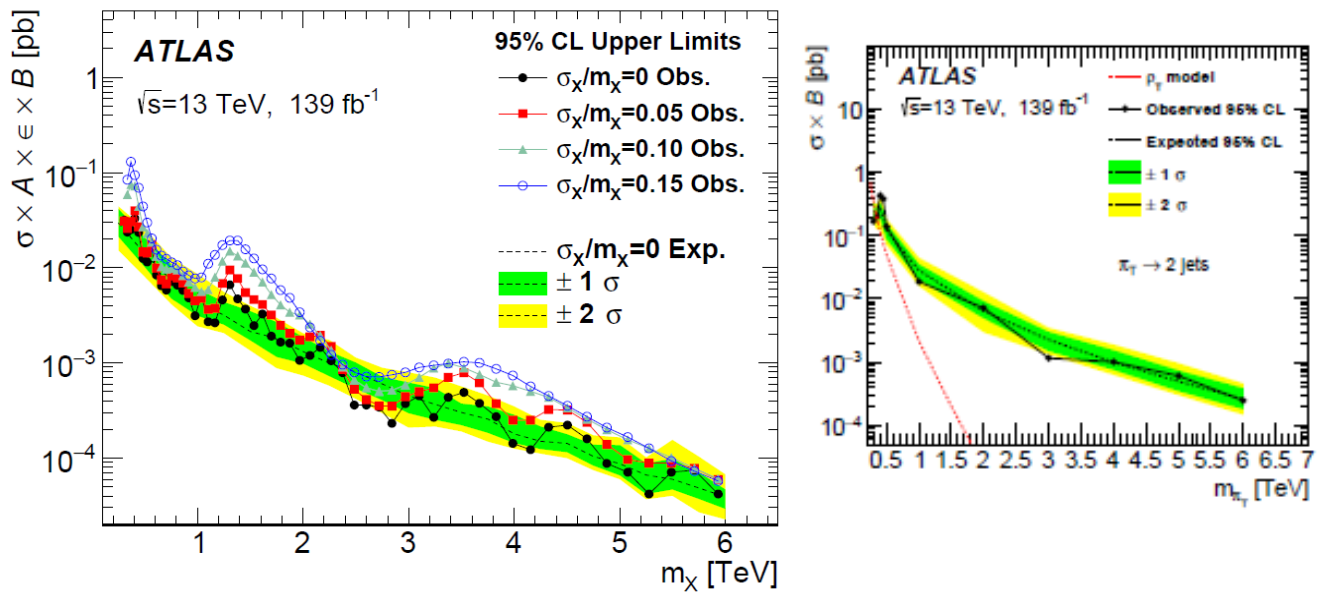

*Figure 6, left: upper limit on the cross-section times acceptance, times BR for an inclusive search of a resonance tagged by a high pT lepton. Right : upper limit for the cross-section times BR for a scalar resonance decaying into two jets.*

Again, this search covers not only ttZ and ttW final states coming from AZ and AW but also from H3+W-, H3+Z and H3+W+ (and their complex conjugate states). One can try to combine the Z particle to the hadronic part to reconstruct exclusive decays of H5 and H5+ resonances. A **VBF select**ion should enhance this signal as mentioned in the previous section.

The excess at 400 GeV in figure 6 (left) appears optimal for a wide resonance assumption. This may be due to a mixture of several contributions with differing masses. The excess of $\pi_T\!\rightarrow\!2$ jets events observed in figure 6 (right) corresponds to a cross-section of a few 100 fb, compatible with what one would expect from the contribution of the parent 5-plet.

### III.3 Alternate interpretation of these excesses

In [12], a Randall Sundrum mechanism, deduced from the AFBb anomaly observed at LEP1, is predicting a possible enhancement of the ttZ coupling. In this interpretation, there is simply a larger cross-section for ttZ but with no significant alteration of the kinematical distributions with respect to the SM. This is radically different from the GM cascade interpretation, where one would expect that, for instance, the **average transverse momentum** distribution of Z/W could be larger than for the SM.

Reference [13] has analysed anomalies for **multi-leptonic states accompanied (or not) by b-jets** and has interpreted them as due to the occurrence of new scalars, H(270) and S(150), with H decaying into S+h(125) or into SS*. Once these masses are fixed, the result depends on one parameter, $\beta^2 g$.

The table below shows a detailed description of these anomalies observed with ATLAS and CMS. The claim is that combining all these discrepancies, there is an **overall 8 s.d. effect**, incompatible with a statistical fluctuation.

This approach is complementary to the one used to establish the presence of the **A(400)** resonance in [2] where, by combining several exclusive channels, one reaches a similar statistical significance **beyond 6 s.d.** If one takes into account the H5(660)->ZZ excess, the ttW excess, one also reaches an overall ~8 s.d. effect. These two observations are clearly independent and reinforce the case for promising discoveries at HL-LHC.

| Selection | Best-fit $\beta_g^2$ | Significance |
|---|---|---|
| ATLAS Run 1 SS $\ell\ell$ and $\ell\ell\ell + b$-jets | $6.51 \pm 2.99$ | $2.37\sigma$ |
| ATLAS Run 1 OS $e\mu + b$-jets | $4.09 \pm 1.37$ | $2.99\sigma$ |
| CMS Run 2 SS $e\mu$, $\mu\mu$ and $\ell\ell\ell + b$-jets | $1.41 \pm 0.80$ | $1.75\sigma$ |
| CMS Run 2 OS $e\mu$ | $2.79 \pm 0.52$ | $5.45\sigma$ |
| CMS Run 2 $\ell\ell\ell + E_{\mathrm{T}}^{\mathrm{miss}}$ $(WZ)$ | $9.70 \pm 3.88$ | $2.36\sigma$ |
| ATLAS Run 2 SS $\ell\ell$ and $\ell\ell\ell + b$-jets | $2.22 \pm 1.19$ | $2.01\sigma$ |
| ATLAS Run 2 OS $e\mu + b$-jets | $5.42 \pm 1.28$ | $4.06\sigma$ |
| ATLAS Run 2 $\ell\ell\ell + E_{\mathrm{T}}^{\mathrm{miss}}$ $(WZ)$ | $9.05 \pm 3.35$ | $2.52\sigma$ |
| Combination | $2.92 \pm 0.35$ | $8.04\sigma$ |

Since then and under the influence of [13], an **exclusive search for S(150)** has been successfully carried out in [14], using the $\gamma\gamma$ and $Z\gamma$ final state, accompanied by missing transverse energy, leptons or b jets. This analysis relies on an exploitation of publicly available data from ATLAS and CMS. The best evidence seems to come from the Etmiss topology which implies that H(270)->SS*, S decaying invisibly, which makes it markedly different from a replica of h(125). Absence of indications into the preferred WW/ZZ final states also makes this particle quite at variance with h(125).

| Type/ Channel | A(tt)Z | A(tt)W | H3+(tb,Wh)W- | H3+(W+h)W+ | H3+(tb)Z |
|---|---|---|---|---|---|
| ATLAS R1 SS ℓℓ and 3ℓ + b | X | X | | X | X |
| ATLAS R1 OS eμ +b | X | X | X | | X |
| CMS R2 SS eμ, μμ, 3ℓ +b | X | X | X | X | X |
| CMS R2 OS eμ | X | X | X | | X |
| CMS R2 3ℓ +ETmiss | X | X | | | X |
| ATLAS R2 SS ℓℓ and 3ℓ + b | X | X | | X | X |
| ATLAS R2 OS eμ + b | X | X | X | | X |
| ATLAS R2 3ℓ +ETmiss | X | X | | | X |

One can ask whether the two approaches **can be distinguished**. GM predicts that these excesses have a **VBF origin,** which offers a discriminating test of the two hypotheses. A proof of the other mechanism relies on the discovery of H(270) into $\gamma\gamma bb$ and $\gamma\gamma\tau\tau$, as recommended in [14]. One can also search for H->S'(96)S(->invisible).

Another discrimination comes from the measurements of ttW and ttZ, which is only affected in the GM case.

The table shows how the various cascades described at the beginning of this section can do the job.

## IV. An extended GM model ?

From the previous sections, one would tend to conclude that all goes well within the GM model. This is not so in the **fermion sector**, as pointed out in the Appendix (cf. III.1):

- How is it that A(400) is accompanied by b jets at a level which allows to tag efficiently A->Zh and A->$\tau\tau$ ?
- How can one understand the ratio BR(A->$\tau\tau$)/BR(A->tt)~0.8% [2] ?

In the following section, I will describe the ingredients needed to resolve these contradictions with the simple GM.

### IV.1 GM+2HDM+ A2HDS

The GM model, in its standard version, predicts universal couplings of the CP-odd Higgs boson A(400) to up, down and leptons, in contradiction with what is suggested by the LHC data (see the Appendix) . In contrast, a 2HDM model allows enhancing the coupling to the leptons and to the b quarks with respect to the coupling to the top quark, which seems requested by the data.

Keeping the triplet structure of the GM model, which prevents violating the $\rho$ parameter constraint, one can add a second doublet which also allows to keep this condition and therefore satisfy our various requirements:

- We create 3 new states, call them H2+- and A2, with bb and ττ couplings enhanced by tβ
- We keep the features of GM which allow to explain the observation of H5->ZZ with BR(ZZ)/BR(WW)=2 and allow for the process H3->hZ

This is not satisfactory since, in doing so, we decrease the coupling to top pairs by tβ$^{-1}$, a problem encountered in [15]. To avoid this, one may use the Aligned-Two-Higgs-Doublet-Scheme mechanism **A2HDS**, a more general scheme invented to suppress FCNC transitions [16], which assumes:

**Y2f=ξfY1f**

where Y1f and Y2f are the Yukawa couplings to the two doublets φ1 and φ2, and where **ξf** is an arbitrary constant which can be complex and differ for top, bottom and lepton.

After diagonalization of the mass matrix, the leptonic coupling of the CP-odd A is multiplied by **ζℓ=(ξℓ−tβ)/(1+ξℓtβ),** instead of -tβ in MSSM. The following table, taken from [17], summarizes the various possibilities. Note that the usual types can be recovered for particular values of ξ. For instance type II is obtained for ξu=infinite and ξd,ℓ=0.

|         | $Y_1^d$ | $Y_1^u$ | $Y_1^\ell$ | $Y_2^d$ | $Y_2^u$ | $Y_2^\ell$ | $\zeta_u$ | $\zeta_d$ | $\zeta_\ell$ |
|---------|---------|---------|------------|---------|---------|------------|-----------|-----------|--------------|
| Type I  | 0 | 0 | 0 | × | × | × | $t_\beta^{-1}$ | $t_\beta^{-1}$ | $t_\beta^{-1}$ |
| Type II | × | 0 | × | 0 | × | 0 | $t_\beta^{-1}$ | $-t_\beta$ | $-t_\beta$ |
| Type X  | 0 | 0 | × | × | × | 0 | $t_\beta^{-1}$ | $t_\beta^{-1}$ | $-t_\beta$ |
| Type Y  | × | 0 | 0 | 0 | × | × | $t_\beta^{-1}$ | $-t_\beta$ | $t_\beta^{-1}$ |
| A2HDS   | × | × | × | × | × | × | $\dfrac{\xi_u - t_\beta}{1 + \xi_u t_\beta}$ | $\dfrac{\xi_d - t_\beta}{1 + \xi_d t_\beta}$ | $\dfrac{\xi_\ell - t_\beta}{1 + \xi_\ell t_\beta}$ |

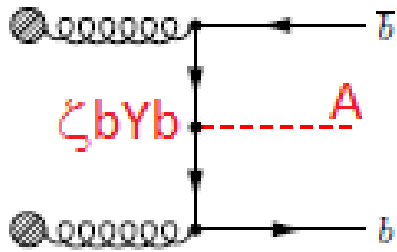

This new sophistication allows interpreting the observed Yukawa constants for the 400 GeV pseudo-scalar, which cannot be understood within MSSM [15]. It also allows increasing the Yukawa coupling of the b quark to this resonance, explaining the observation of events tagged by b jets. Even assuming that tβ ~1, one can enhance the Yℓ coupling by ζℓ=-25 with respect to the SM by taking ξℓ~-1+0.08, to satisfy experimental observations. The same could be done for ζb, while one can keep ζt close to 1, as suggested by the data.

In this extended GM scheme, the physical state A(400) can be a mixture of H3 and A2, and only the latter component will be enhanced, reducing effect by a mixing coefficient. This can be compensated by increasing |ζℓ| above 25. In the same way the coupling of A(400) to hZ will be decreased by a mixing factor, which, again, can be compensated by changing the mixing angles α and θH in the GM model.

Note that this extension implies an extra charged boson from a doublet, H2+, in addition to the GM states, H3+ and H5+. One also expects H2, the heavy CP-even component of the two doublet part.

Adding new isosinglets is also possible, which may be needed to accommodate h(96) and the newly observed S(151).

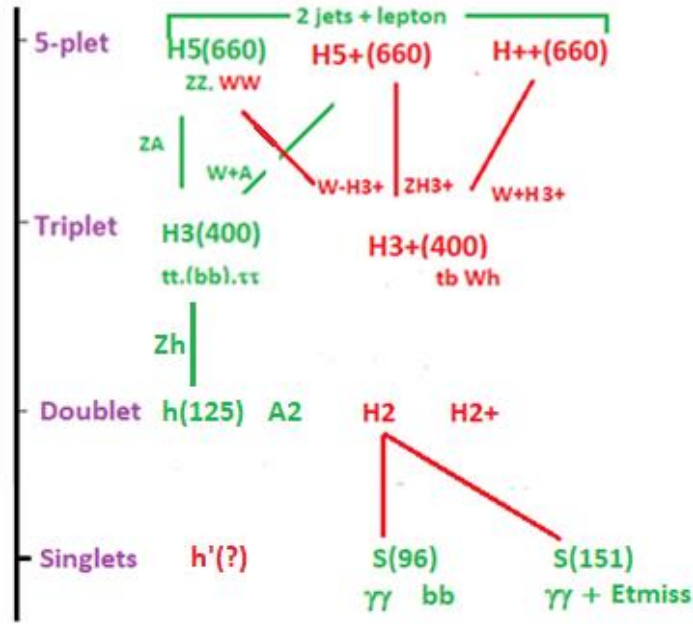

To conclude there is a good prospect that the GM could be extended to satisfy the experimental requirements suggested by LHC indications. A very rich spectrum of scalars is therefore expected, with eleven new states, four of them already appearing in the data. Above picture summarizes the spectrum expected in an extended GM model, EGM.

## IV.2 g-2

Another reason to extend the GM model is provided by a recent interpretation [17] of the **(g-2)μ anomaly** in terms of the two-loop diagram shown below. This diagram is a priori second order but turns out to be dominant, given the **kinematic suppressions** occurring at the tree level. In the 1-loop diagrams, one has to pay the price of two small Yukawa Yμ couplings, and in addition pay another mμ factor for the muon chirality flip. In the 2-loop diagram, Yμ, appears only once, and the same coupling also takes care of the chirality flip.

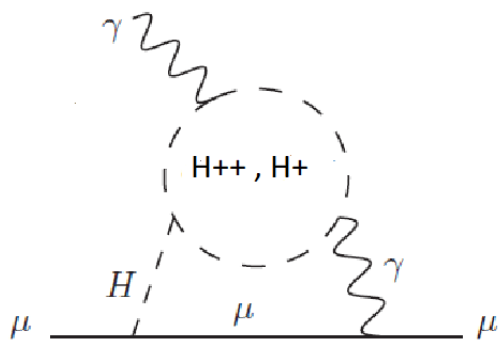

[17] claims, rightly, that it is conceivable that this type of diagram can dominantly contribute to (g-2)μ and explain the anomaly, provided that there is a neutral Higgs H which strongly couples to μμ and to H++ H- -. This can happen even if the exchanged Higgs bosons weigh several 100 GeV, as is the case in our scenario.

While this does not happen in the minimal GM model, where the Higgs scalar H5 does not couple to fermions, it could happen in the extended version described above where one has a H2 particle which could be not too heavy, with large couplings to leptons.

Finally, one needs to concoct the **Higgs potential** such that there is a large coupling λ of H to H++, as was done in [17].

Note also that there is a factor four enhancement in this (g-2)μ contribution due to the two couplings of H++ to photons.

Admittedly, this discussion is only qualitative and a fully worked out model is still needed, which goes beyond the scope of the present paper.

The lesson to retain is that one needs to extend the GM set up by an additional doublet with two potential benefits:

- Understanding the origin of the X(400)->ττ signal and the b-tagging enhancement
- Providing an interpretation of the (g-2)μ anomaly in terms of the heavy scalars indicated by LHC

## V.      Future prospects at LHC

From previous sections, one expects LHC:

- To confirm the existence of A(400) through top pairs, Zh and ττ +b
- To confirm H5(660) into ZZ/WW through VBF
- To confirm indications for a charged Higgs into ZW through VBF
- To search for h'->2h which is expected to be the dominant decay mode (figure 10)
- To reconstruct H5, H5+ using ttW and ttZ final states
- To discover H3+ by using hadronic final states tagged by a high pT leptons from Z/W decays
- To search for H5++

At LHC, H5++ can be singly produced through VBF. It decay modes are into W+W+ and H3+W+ (and complex conjugate), the latter being predominant with H3+ decaying into tb or hW+. This gives topologies of the type tbW->bbW+W+ and hW+W+->bbW+W+, which can be searched for by selecting the VBF mode. [18] gives a pessimistic prediction for this channel as shown in figure 7.

To fully assess GM, searching for h' is a priority. From the first table of section II.3, one concludes that:

- ggF->h' cross-section is reduced by a factor 30 with respect to a SM Higgs
- VBF remains the best prospect since it is only reduced by a factor 2 with respect to SM
- For mh'>2mh(125) the dominant decay mode is 2h (see Appendix)

For mh'=600 GeV, the predicted VBF cross-section is ~100 fb plus ~50 fb from ggF, with a dominant decay into 2h (see Appendix for the present experimental indications) .

Finally, not to be forgotten, is the clarification of the status of h(96), only observed by CMS. h(96) seems unrelated to the standard GM model but could be accommodated in and extended version as an extra isosinglet. It could also be interpreted as the RS scalar called **Radion**. As stated in [19], this particle will, through mixing effects, influence the properties of h(125).

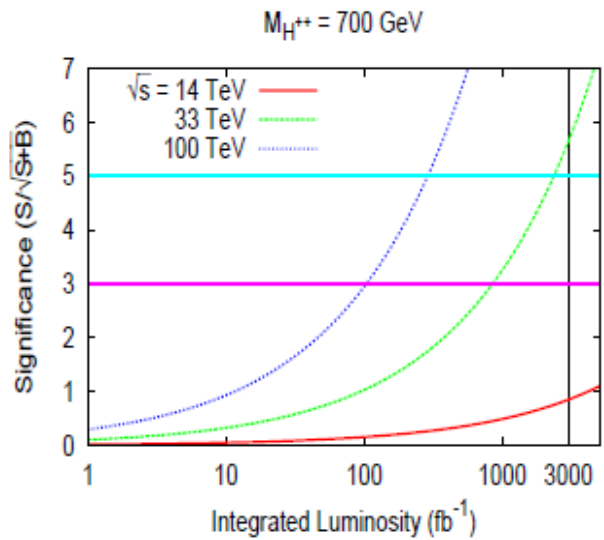

*Figure 7: Expected significance for a H++(700) vs. the integrated luminosity delivered by LHC at 14 TeV*

# VI.    Future prospects at e+e- colliders

## VI.1 Expected rates at a TeV e+e- collider

Figure 8 shows a clear advantage in reaching 1 TeV at a future e+e- collider. All GM final states could then be covered with the exception of doubly charged scalars requiring ECM> 1.4 TeV. H5++ can be singly produced in association with W-W- but the corresponding cross-section falls below 0.1 fb, as predicted in [20].

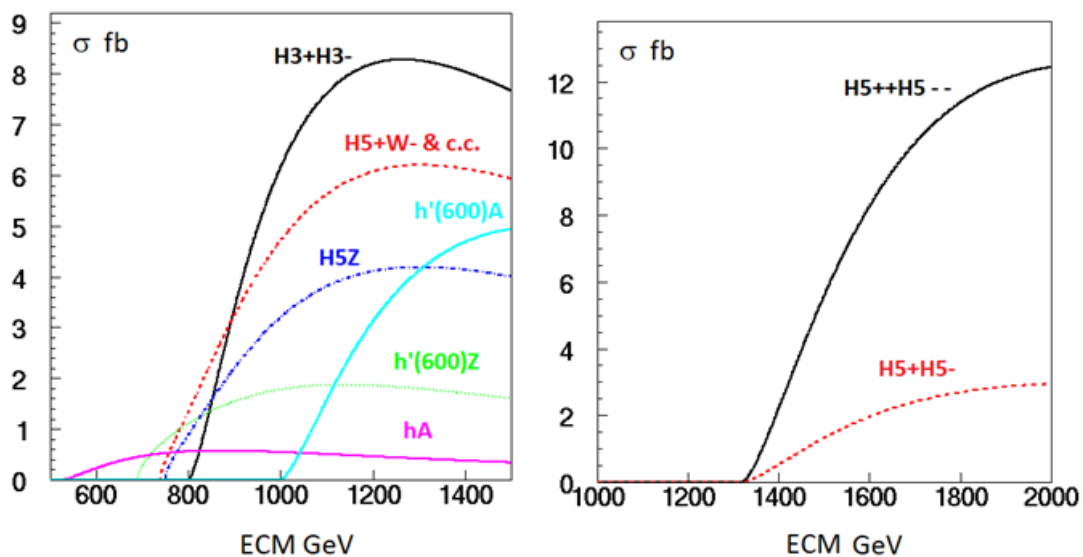

*Figure 8: Predicted cross-sections for various GM channels at ILC versus the centre of mass energy ECM in GeV.*

## VI.2 Mass, width and cross-section measurements

For H5Z and h'Z channels, one can proceed as for h(125)Z, using Z into lepton pairs. This method will also give access to the total width and the invisible width. Optimal centre of mass energy is ~mh'+200 GeV for h'Z and ~1000 GeV for H5Z.

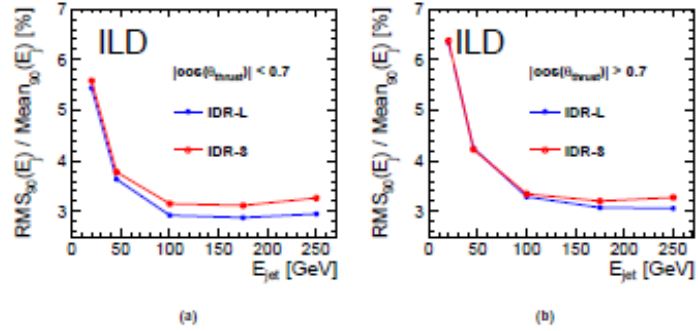

*Figure 9: Expected jet resolution at ILD versus jet energy for two versions of the detector. Plot (a) corresponds to the barrel region, while plot b is for the end-cap.*

The situation is however quite different from the SM measurements for h(125)Z since there is a negligible ZZ and h(125)Z background for these large masses. The recoil mass width comes from the large width of these heavy resonances, meaning that one can also use hadronic decays of W/Z/h (the later for hA) without a significant degradation. One can further improve the reconstruction accuracy by imposing the Z/W/h mass constraint.

Since H5+ and H5 are **mass-degenerate**, the recoil mass technique is delicate for H5Z and H5+W-modes. One needs to use the ability of the detector to separate W and Z masses. This ability has been thoroughly studied [21] and works quite well with the resolutions of figure 9.

If h' and H5 are mass-degenerate, a discrimination is still possible noticing that they share different final states, h' decaying dominantly into h(125)h(125).

The table below gives an estimate of the expected accuracies for resonances produced in e+e- at ECM=1 TeV with an integrated luminosity of 8000 fb-1. This assumption is just a linear extrapolation of the integrated luminosity assumed for ILC at 500 GeV: 4000 fb-1. Detailed studies are obviously needed to assess this figure, which crucially depends on the limits on **power consumption**. The tantalizing prospect of an ERL usage of ILC could further improve these figures [22].

| Mode | $\delta M$ MeV | $\delta\sigma/\sigma$ % | $\delta\Gamma/\Gamma$ % | $\delta\Gamma inv/\Gamma$ |
|------|------|------|------|------|
| H5Z | 280 MeV | 0.7% | 0.5% | 0.02% |
| h'(600)Z | 180 MeV | 1% | 0.7% | 0.03% |
| A(400)h(125) | 460 MeV | 2.4% | 1.7% | 0.2% |

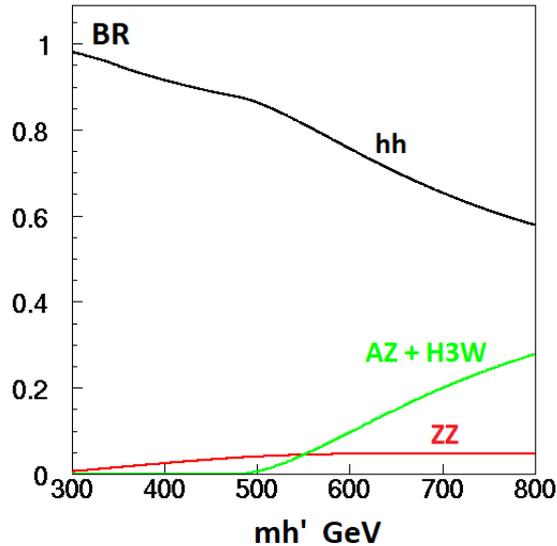

*Figure 10: Branching ratio of the main decay modes of h' versus the mass of this particle.*

Clearly these estimates are too optimistic since they assume a perfect W-Z separation in hadronic modes, while one knows that, due to b semi-leptonic decays, Z->bb can leak into W. One can veto against b quarks with no damage for W decays. A full study with a **realistic simulation** is therefore required. Presently, my intention is to draw the attention on the importance of a **precise particle flow reconstruction** for Z/W/h hadronic decay in the detectors planned for future e+e-colliders and give a crude estimate of the performances that one can expect from an ILD-like detector.

**Reconstruction efficiencies** could be low, given the complexity of such events, comprising typically 8 jets. The efficiency for reconstructing n jets goes, roughly, like $\Omega^n$, where $\Omega$ is the solid angle covered by the detector and n is the number of jets. For instance b-tagging efficiencies tend to be low for $|\cos\theta|<0.8$ which means that less than 20% of these events would be fully b-tagged.

For the channel H3+H3-, the mass can be deduced by measuring the cross-section, given its large cross section and steep dependence with the centre of mass energy.

Figure 10 shows the variation of the BR of h' for the main final states. Details of its derivation can be found in the Appendix. It shows that this particle is best identified in the hh mode. A 1 TeV linear collider should cover this search up to a mass of ~900 GeV in the h'Z channel. Note that, at 1 TeV, the SM background Zhh is an order of magnitude lower than the cross-section for Zh'(600), which allows a very clean measurement of the h' properties.

In conclusion, these examples show how a LC operating up to 1 TeV will be an insuperable instrument to perform the critical measurements needed to disentangle the very rich spectroscopy expected from the GM model.

## Conclusions and future prospects

The lesson of these investigations is that one should remain very open-minded in our interpretations of the LHC data and that light objects are not yet excluded.

This note has attempted to describe and interpret the great diversity of indications for BSM physics, observed by ATLAS and CMS. Two entirely different approaches, one based on **spectroscopy**, the other on **topology** (multileptonic events associated to b jets, ttW), lead to the conclusion that a **BSM mechanism** could be present in LHC data. Provided that phenomenology can provide a **consistent interpretation** of these apparently unrelated evidences, these effects should be taken seriously and drive the **search strategy** at LHC.

The conclusion of this paper is that they agree quite well with the **Georgi-Machacek** framework, with the exception of the couplings of A(400) to b and τ fermions which seems incompatible to the coupling to top quarks. As pointed out in section IV, a solution to this problem can be found with a reasonable extension of the GM model. With this **extension**, one could expect a significant contribution of these scalars to the **(g-2)μ** anomaly.

Using the LHC measurements, one can precisely determine the parameters of this model and forge a useful tool to predict what is left to be measured or discovered both at LHC and at future e+e- machines. These results allow to design a proper strategy for discoveries at LHC avoiding the usual criticisms against open minded searches: the so called "**look elsewhere**" criteria, which inexorably weaken evidences, can be eliminated given that all masses and most couplings are by now determined, with no arbitrariness in the choice of channels. This reminds us the situation when LHC was starting its search for the SM Higgs boson.

A major virtue of these findings is avoiding the ongoing **blind race for heavier masses**, with the implicit statement that SUSY is just behind the corner, at the risk of sacrificing genuine signals. One can already witness this dangerous tendency by noticing that, with more analysed luminosity, one does not observe progress for relevant low mass regions. This is true for the searches for A(400) and h'.

With an appropriate approach for discoveries at LHC, one will not only be able to confirm/discard such effects but also to design an optimum strategy for improving present searches. A good example of this is the **"cascade" approach** where one tags by a high pT lepton an inclusive search for hadronic decays. One can allow further selections for this search (e.g. b-tag tagging) and try to reconstruct the mass of the parent 5-plet heavy resonance.

The recent evidence for an additional scalar, **S(151)**, observed into two photons mostly accompanied by missing transverse energy, offers an additional source of "cascades". It could presumably be accommodated as an isosinglet in a further extension of the GM model, similarly to the extensions proposed for the N2HDM models. Indications for a scalar at 96 GeV could be integrated in the same way.

This is also true for the **ttW analysis** where one could try to reconstruct exclusively the resonance producing the 50% excess. Even at the inclusive level, one can observe a more striking deviation originating from H5+, then all these events come from ZW fusion, while the SM events originate mostly from ggF. This feature is a **critical prediction of my analysis**, which should be relatively easy to verify with standard methods. In this way, one would isolate a very pure GM signal and, ultimately, be able to reconstruct the mass of the parent Higgs from the 5-plet. This example beautifully illustrates the importance of **VBF physics** and the request that the detectors can still perform well on forward jet tagging in the HL-LHC phase.

A straightforward issue is clearly to consolidate the **4 leptons analysis for the ZZ** final state and combine the two LHC experiments. Separating ggF from VBF is of crucial importance.

An ultimate proof of the GM model – its **smoking gun** - is to observe a **charged scalar** H5+->ZW+, not allowed in MSSM. The other decay modes are AW+ and H3+Z, not easy to reconstruct. H3+ can decay into tb (see section V of the Appendix) or into h(125)W+.

For what concerns **future e+e- HE colliders,** like ILC and CLIC, these indications, when confirmed, will support the need to reach at least 1 TeV, with the highest possible luminosity. A large fraction of the program − h(125)A(400), H5(660)Z, H3+H3-, h'Z − can be performed with energies **reaching ~1 TeV**, while extending this energy to 1.5 TeV, one can observe the scalars with **double charge**. Final states from the 5-plet are **very complex** and will require not only high **luminosity** but also an **almost perfect detector** with best possible angular coverage for jet reconstruction and flavour tagging.

To conclude, there is a fascinating possibility that **a simple extension of a** model, proposed in 1985, could provide a plausible interpretation of the large set of anomalies observed by ATLAS and CMS.

**Acknowledgements**: *I am grateful to my colleagues from IJCLab M. Davier, A. Falkowski, R. Poeschl and Z. Zhang for kindly encouraging this work. Adam Falkowski's invaluable help is gratefully acknowledged for clarifying the various issues concerning the (g-2)μ contributions of heavy scalars. George Moultaka's expertise on the delicate issues of custodial symmetry in models with triplets deserves my gratitude. I thank Tanmoy Modak, from Natl. Taiwan U., for patiently discussing the issue of CPV in the scalar and pseudo-scalar sectors. I also thank Carlos Wagner for his interest and for pointing his work [5] in the interpretation of the ATLAS indication for A(400)->hZ.*

**References:**
*[1] Indications for extra scalars at LHC? -- BSM physics at future e+e− colliders*
*François Richard (IJCLab, Orsay). Jan 14, 2020.*
*e-Print: arXiv:2001.04770*
*[2] Evidences for a pseudo scalar resonance at 400 GeV. Possible interpretations*
*François Richard (IJCLab, Orsay) (Mar 16, 2020)*
*e-Print: 2003.07112*
*[3] DOUBLY CHARGED HIGGS BOSONS*
*Howard Georgi (Harvard U.), Marie Machacek (Northeastern U.)(Jun 1, 1985)*
*Published in Nucl.Phys.B 262 (1985) 463-477*
*[4] Higgs Boson Triplets With MW = MZ cosϑω*
*Michael S. Chanowitz (LBL, Berkeley), Mitchell Golden (LBL, Berkeley) (Sep 1, 1985)*
*Published in: Phys.Lett.B 165 (1985) 105-108*
*[5] Wrong sign bottom Yukawa coupling in low energy supersymmetry*
*Nina M. Coyle(Chicago U. and Chicago U., EFI), Bing Li (Chicago U. and Chicago U., EFI), Carlos E.M. Wagner (Chicago U. and Chicago U., EFI and Chicago U., KICP and Argonne) (Feb 25, 2018)*
*Published in: Phys.Rev.D 97 (2018) 11, 115028*
*e-Print: 1802.09122*

[6] Measurements of differential cross-sections in four-lepton events in 13 TeV proton-proton collisions with the ATLAS detector
ATLAS Collaboration Georges Aad (Marseille, CPPM) et al. (Mar 2, 2021)
e-Print: 2103.01918
Search for heavy resonances decaying into a pair of Z bosons in the ℓ+ℓ−ℓ′+ℓ′−\ell^+\ell^-\ell'^+\ell'^-ℓ+ℓ−ℓ′+ℓ′− and ℓ+ℓ−νν¯\ell^+\ell^-\nu\bar\nuℓ+ℓ−νν¯ final states using 139 fb−1 of proton-proton collisions at \sqrt{s}=13 TeV with the ATLAS detector
ATLAS Collaboration Georges Aad (Marseille, CPPM) et al. (Sep 30, 2020)
e-Print: 2009.14791

[7] Search for charged Higgs bosons produced in vector boson fusion processes and decaying into a pair of W and Z bosons using proton-proton collisions at sqrt(s) = 13 TeV
CMS Collaboration Albert M Sirunyan (Yerevan Phys. Inst.) et al. (May 8, 2017)
Published in: Phys.Rev.Lett. 119 (2017) 14, 141802
e-Print: 1705.02942

[8] Search for resonant WZ production in the fully leptonic final state in proton-proton collisions at \sqrt{s} = 13 TeV with the ATLAS detector
ATLAS Collaboration M. Aaboud et al. (Jun 5, 2018)
Published in: Phys.Lett.B 787 (2018) 68-88
e-Print: 1806.01532

[9] Triplet Higgs boson collider phenomenology after the LHC
Christoph Englert (Durham U., IPPP) , Emanuele Re (Oxford U., Theor. Phys.), Michael Spannowsky (Durham U., IPPP) (Feb 26, 2013)
Published in: Phys.Rev.D 87 (2013) 9, 095014
e-Print: 1302.6505

[10] Measurement of the Higgs boson production rate in association with top quarks in final states with electrons, muons, and hadronically decaying tau leptons at \sqrt{s} = 13 TeV
CMS Collaboration
Albert M Sirunyan (Yerevan Phys. Inst.) et al. (Nov 6, 2020)
e-Print: 2011.03652

[11] Search for dijet resonances in events with an isolated charged lepton using \sqrt{s}=13 TeV proton-proton collision data collected by the ATLAS detector
ATLAS Collaboration Georges Aad (Marseille, CPPM)et al.(Feb 26, 2020) Published in: JHEP 06 (2020) 151
e-Print: 2002.11325

[12] Can LHC observe an anomaly in ttZ production?
François Richard (Orsay, LAL)(Apr 12, 2013)
e-Print: 1304.3594

[13] The emergence of multi-lepton anomalies at the LHC and their compatibility with new physics at the EW scale
Stefan Buddenbrock (U. Witwatersrand, Johannesburg, Sch. Phys.), Alan S. Cornell (Witwatersrand U.), Yaquan Fang (Beijing, Inst. High Energy Phys.), Abdualazem Fadol Mohammed (Witwatersrand U. and Beijing, Inst. High Energy Phys. and Beijing, GUCAS), Mukesh Kumar (Witwatersrand U.) et al. (Jan 16, 2019)
Published in: JHEP 10 (2019) 157
e-Print: 1901.05300

[14] Accumulating Evidence for the Associate Production of a Neutral Scalar with Mass around 151 GeV
Andreas Crivellin (Zurich U. and PSI, Villigen and CERN, Yaquan Fang (Beijing, Inst. High Energy Phys. and Beijing, GUCAS), Oliver Fischer (Liverpool U.),  Abhaya Kumar (U. Witwatersrand, Johannesburg, Sch. Phys. and iThemba LABS), Mukesh Kumar et al. (Sep 6, 2021)
e-Print 2109.02650

[15]Possible indications for new Higgs bosons in the reach of the LHC: N2HDM and NMSSM interpretations
Thomas Biekötter (DESY), Alexander Grohsjean (DESY), Sven Heinemeyer (Madrid, IFT and Madrid, Autonoma U. and Madrid, IFT), Christian Schwanenberger (DESY and Hamburg U.), Georg Weiglein (DESY and Hamburg U.) (Sep 2, 2021)
e-Print: 2109.01128

[16] Yukawa Alignment in the Two-Higgs-Doublet Model
Antonio Pich (Valencia U. and Valencia U., IFIC), Paula Tuzon(Valencia U. and Valencia U., IFIC) (Aug, 2009)
Published in: Phys.Rev.D 80 (2009) 091702
e-Print: 0908.1554

 [17] Muon g−2 in two-higgs-doublet model with type-II seesaw mechanism
Chuan-Hung Chen (Taiwan, Natl. Cheng Kung U. and NCTS, Taipei and Taiwan, Natl. Taiwan U.), Cheng-Wei Chiang(NCTS, Taipei and Taiwan, Natl. Taiwan U.), Takaaki Nomura(Sichuan U.)(Apr 7, 2021)
e-Print: 2104.03275

[18] Search for doubly charged Higgs bosons through vector boson fusion at the LHC and beyond
G. Bambhaniya (Ahmedabad, Phys. Res. Lab), J. Chakrabortty (Indian Inst. Tech., Kanpur), J. Gluza(Silesia U.), T. Jelinski (Silesia U.), R. Szafron (Alberta U.)(Apr 15, 2015)
Published in: Phys.Rev.D 92 (2015) 1, 015016
e-Print: 1504.03999

[19] Search for a light radion at HL-LHC and ILC250
Francois Richard (Orsay, LAL)(Dec 18, 2017)
e-Print: 1712.06410

[20]Phenomenology of the Georgi-Machacek model at future electron-positron colliders
Cheng-Wei Chiang (NCTS, Hsinchu and Taiwan, Inst. Phys. and Taiwan, Natl. Central U.), Shinya Kanemura (Toyama U.), Kei Yagyu(U. Southampton (main)) (Oct 21, 2015)
Published in: Phys.Rev.D 93 (2016) 5, 055002
e-Print: 1510.06297

[21]International Large Detector: Interim Design Report
ILD Concept Group
Halina Abramowicz (Tel Aviv U.) et al. (Mar 2, 2020)
e-Print: 2003.01116 [physics.ins-det]

[22] A high luminosity superconducting twin e+e−e^+e^-e+e− linear collider with energy recovery
V.I. Telnov (Novosibirsk, IYF and Novosibirsk State U.) (May 23, 2021)
e-Print: 2105.11015

[23] Vacuum stability conditions for Higgs potentials with SU(2)LSU(2)_LSU(2)L triplets
Gilbert Moultaka (U. Montpellier, L2C), Michel C. Peyranère (U. Montpellier 2, LUPM) (Dec 27, 2020)
Published in: Phys.Rev.D 103 (2021) 11, 11500
e-Print: 2012.13947

[24]Scalar decays to γγ\gamma\gammaγγ, ZγZ\gammaZγ, and WγW\gammaWγ in the Georgi-Machacek model
Celine Degrande(CERN), Katy Hartling (Ottawa Carleton Inst. Phys.), Heather E. Logan (Ottawa Carleton Inst. Phys.) (Aug 29, 2017)
Published in: Phys.Rev.D 96 (2017) 7, 075013, Phys.Rev.D 98 (2018) 1, 019901 (erratum)
Print: 1708.08753

[25]Naturalness problems for rho = 1 and other large one loop effects for a standard model Higgs sector containing triplet fields
J.F. Gunion (UC, Davis), R. Vega (UC, Davis), J. Wudka (UC, Davis)(Sep, 1990)
Published in: Phys.Rev.D 43 (1991) 2322-2336

[26] *Unitarity bounds in the Higgs model including triplet fields with custodial symmetry*
*Mayumi Aoki (Tokyo U., ICRR), Shinya Kanemura (Toyama U.) (Dec 27, 2007)*
*Published in: Phys.Rev.D 77 (2008) 9, 095009, Phys.Rev.D 89 (2014) 5, 059902 (erratum)*
*e-Print: 0712.4053*

[27] *Radiative corrections to the Z b anti-b vertex and constraints on extended Higgs sectors*
*Howard E. Haber (UC, Santa Cruz), Heather E. Logan (UC, Santa Cruz) (Aug, 1999)*
*Published in: Phys.Rev.D 62 (2000) 015011*
*e-Print: hep-ph/9909335*

[28] *Observation of electroweak production of same-sign W boson pairs in the two jet and two same-sign lepton final state in proton-proton collisions at \sqrt{s} = 13 TeV*
*CMS Collaboration Albert M Sirunyan (Yerevan Phys. Inst.) et al. (Sep 18, 2017)*
*Published in: Phys.Rev.Lett. 120 (2018) 8, 081801*
*e-Print: 1709.05822*

[29] *M. Zaro and H. Logan, "Recommendations for the interpretation of LHC searches for $H_{05}$, $H_5$, and $H_5$ in vector boson fusion with decays to vector boson pairs", CERN Report LHCHXSWG-2015-001, 2015.*

[30] *Double Higgs production at LHC, see-saw type II and Georgi-Machacek model*
*S.I. Godunov(Moscow, ITEP and Novosibirsk State U.), M.I. Vysotsky (Moscow, ITEP and Moscow, MIPT and Moscow Phys. Eng. Inst.), E.V. Zhemchugov (Moscow, ITEP and Moscow Phys. Eng. Inst.) (Aug 1, 2014)*
*Published in: J.Exp.Theor.Phys. 120 (2015) 3, 369-375*
*e-Print: 1408.0184*

[31] *Higgs boson pair productions in the Georgi-Machacek model at the LHC*
*Jung Chang (NCTS, Hsinchu), Chuan-Ren Chen (Taiwan, Natl. Normal U.), Cheng-Wei Chiang (Taiwan, Natl. Taiwan U. and NCTS, Hsinchu and Taiwan, Inst. Phys.) (Jan 23, 2017)*
*Published in: JHEP 03 (2017) 137*
*e-Print: 1701.06291*

[32] *Search for resonant and non-resonant Higgs boson pair production in the bb̄ τ+τ− decay channel using 13 TeV pp collision data from the ATLAS detector*
*ATLAS Collaboration July 29 2021*
*ATLAS-CONF-2021-030*

[33] *Search for resonant Higgs boson pair production in four b quark final state using large-area jets in proton-proton collisions at 13 TeV*
*CMS Collaboration July 27 2021*
*CMS-PAS-B2G-20-004*

[34] *Search for resonances decaying into photon pairs in 139 fb−1^{-1}−1 of pp collisions at s=\sqrt{s} =s= 13 TeV with the ATLAS detector*
*ATLAS Collaboration*
*Georges Aad(Marseille, CPPM) et al. (Feb 26, 2021)*
*e-Print: 2102.13405*

[35] *Can the 750-GeV diphoton resonance be the singlet Higgs boson of custodial Higgs triplet model?*
*Cheng-Wei Chiang(Taiwan, Inst. Phys. and Taiwan, Natl. Central U. and NCTS, Hsinchu), An-Li Kuo(Taiwan, Natl. Central U.) (Jan 24, 2016)*
*Published in: Phys.Lett.B 760 (2016) 634-640*
*e-Print: 1601.06394*

# APPENDIX

## I.     The Georgi-Machacek model for pedestrians

This model allows having an exotic spectrum of **scalars with double charges**, without violating the so-called **ρ parameter constraint**. At the tree level, one has:

$$\rho = \frac{m_w^2}{m_z^2 \cos^2 \vartheta_w} = \frac{\sum_i v_i^2 [T_i(T_i + 1) - Y_i^2]}{\sum_i 2Y_i^2 v_i^2}$$

In GM, one has two triplets, $T_1 = T_2 = 1$, with the same vacuum expectations, $v_1 = v_2$. The first triplet is real and has $Y_1 = 0$, the second is purely imaginary and has $Y_2 = 2$. Therefore $\rho = 0$, at the tree level. At the loop level, the Higgs potential is complex, and allows extra loop contributions to the $\rho$ parameter, which introduces an additional source of uncertainties[4] with respect to the SM.

By having higher isospin representations, GM allows to have a **direct coupling of H+ to WZ**, which in the usual MSSM case is forbidden at the tree level.

One has the following particle content:

A 5-plet   **H5--   H5-   H5   H5+   H5++** with mass degeneracy.

These particles are fermiophobic and can only be produced by VBF and not by ggF, through a fermionic loop. H5+ can couples to ZW.

A triplet   **H3-   CP-odd H3 (called A)   H3+** with mass degeneracy.

H3+ does not couple to ZW but couples to fermions and to hW.

Two singlets called **h(125) and h' ,** which can mix with a mixing angle $\alpha$.

Below are summarized the main LHC observations and the GM interpretations.

H5 is observed in ZZ at 660 GeV, by selecting four leptons and combining, unofficially [1] CMS and ATLAS data. ATLAS, with the full statistics, confirms these findings and, in addition, shows an indication for VBF->ZZ. No signal is observed into WW. GM interpretation of this behaviour: in HWW/HZZ~2 in MSSM/SM, while HWW/HZZ~0.5 in GM.

H5, H5+, H5++ can couple to the Higgs triplet, with the following decay modes:

- H5->AZ, explaining the ~20% excess in ttZ indicated by ATLAS and CMS
- H5+->AW+, explaining the ~50% excess in ttW observed by ATLAS and CMS
- H5+->ZW hint in ATLAS and CMS at ~500 GeV

H5++->W+W+ and W+H3+, the later giving ZW+W+ and tbW+. A limit is set on W+W+ by CMS.

These channels contribute to an inclusive search in 2 jets with lepton tag (from W/Z) in ATLAS, which gives an indication around 400 GeV.

---

[4] This aspect has been brought to my attention by G. Moultaka, see [23] and references therin, for a thorough discussion of these models.

Triplets can couple to a vector boson+h:

- A->Zh(125) observed in μμ/ee +bjet-tag in ATLAS
- H3+->W+h(125) not observed

and to heavy fermions:

- A->tt in CMS, ττ +bjet-tag in ATLAS
- H3+->tb, τ+ν not observed

The two singlets h(125) and h'(?) have the usual decays:

- h(125)->ZZ*/WW*, γγ, ττ, μμ, bb, cc, gg
- h'(?)->WW/ZZ, hh, tt, ττ, μμ, bb, cc, gg
- h', if heavy enough, goes into AZ, H3+W-

## II.    Quantitative treatment of the GM model

In this model, the various relevant couplings depend on the following parameters:

- Two vacuum expectations $v\chi$ and $v\phi$, which are related to the SM vacuum expectation v by the formula $v^2=8v\chi^2+v\phi^2$, v~246 GeV.
- Two mixing angles $\theta_H$ and $\alpha$, with $\cos\theta_H=c_H=v\phi/v$ and $\sin\alpha=s\alpha$

The various couplings to bosons and fermions are given in terms of these parameters as shown in the table of section II. I have used primarily [24] and [25] to derive these couplings.

### II.1  h(125) constraints

The present conclusion seems to be that the GM model, as it is, gives a qualitative description of the various indications observed at LHC in the bosonic decays. At the quantitative level, it allows to fulfil the constraints imposed by Higgs precision measurements, if the two mixing angles fall within an interval defined in figure 11.

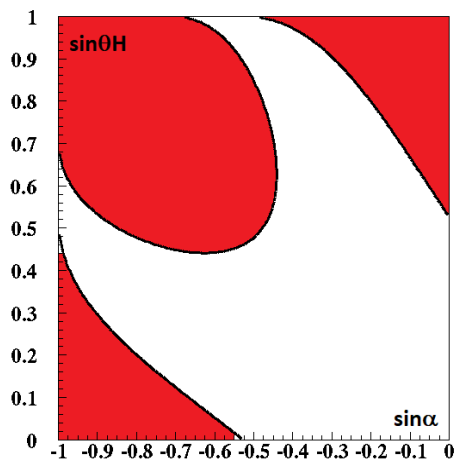

*Figure 11: Allowed region for the two mixing angles of the GM.*
*The two contours correspond to h(125)ZZ/SM (amplitude) between 0.85 and 1.15.*

## II.2 Unitarity bounds on GM scalar masses

These bounds were derived by [26]. Lets briefly recall these results which can be read from figure 12 :

- mH3<400 GeV
- mH5<650 GeV
- mh'<700 GeV

Precision measurements from LEP1 on Zbb predict that **mH5~√3mH3**, in **striking agreement** with LHC present indications. Zbb also provides a constraint on $\tan\theta_H$ as shown in figure 13 from [27].

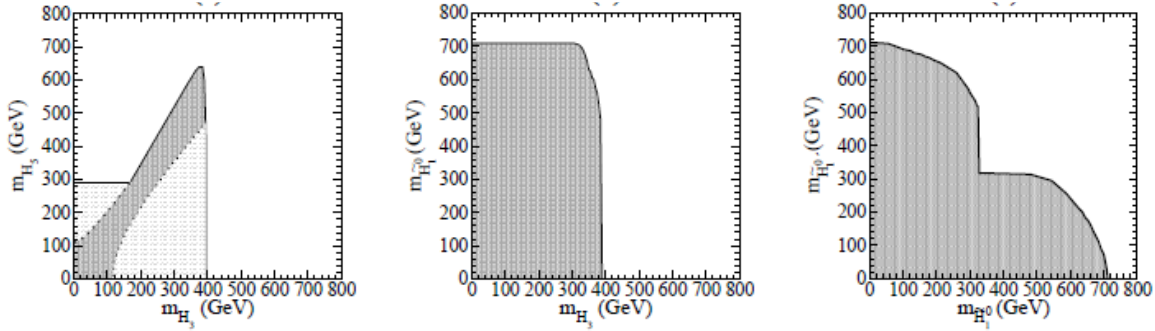

*Figure 12: Allowed regions of the masses of the GM Higgs bosons. In figure (a), the light shadowed regions are excluded by Zbb. H01 is the SM boson while H01' is the other singlet.*

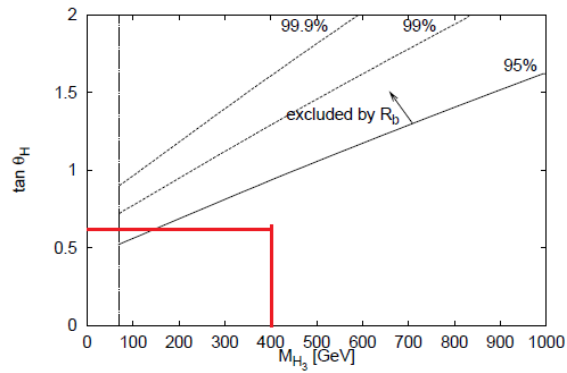

*Figure 13: Bounds from Rb for the G-M model with SU(2)c symmetry. Indicated by red coordinates is my solution.*

## III.  Constraints from LHC indications
### III.1 Fermion couplings

Recall ATLAS and CMS observations:

- Associated production of A+bjet for hZ and $\tau\tau$ channels
- Yukawa coupling YAtt between 0.3 and 0.7

These features requires an extended GM model as discussed in section IV.1. Using the notations of this section, SM Yukawa couplings are multiplied by $\zeta f$.

Figure 14 shows the variation of $\zeta\tau$ vs $\zeta b$ for 3 values of $\zeta t$: 0.3, 0.5 and 0.7. One sees that for a given value of $\zeta b$, it is possible to adjust $\zeta\tau$ to reproduce the ATLAS data. It is however fair to say that small values of $\zeta b$ require unnaturally high values of $\zeta\tau$. Therefore, the most natural choice is $\zeta b$~20. The red line shows the MSSM relation between $\zeta b$ and $\zeta\tau$, which leads to solution $\zeta b$~20. In MSSM this reads as tanβ=20, which may sound acceptable, unless one notices that this result is **incompatible** with the standard relation ζt=1/tanβ. In the model of [16], one can maintain ζt close to 1.

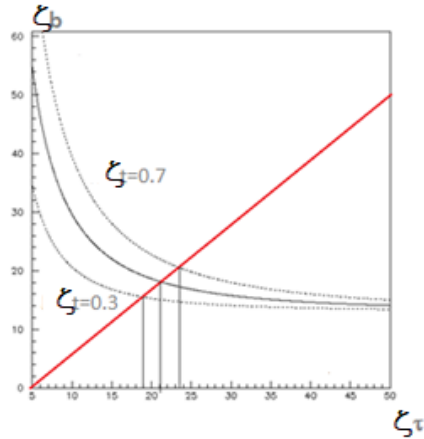

*Figure 14: Using ATLAS indications for a 400 GeV resonance into tau pairs + b quarks and the CMS analysis for A(400) into top pairs, this figure shows the solutions for A couplings to tau and b for the ζAtt interval deduced from CMS. The red line assumes the MSSM relation between tau and b couplings.*

For what concerns GM, one expects ζt=ζb=tanθ_H~0.6, clearly also incompatible with LHC observations.

### III.3 Bosonic constraints

Three constraints were used:

- The h(125)ZZ coupling constraint from LHC already discussed
- The H5(660)->ZZ constraint
- The A(400)->h(125)Z constraint

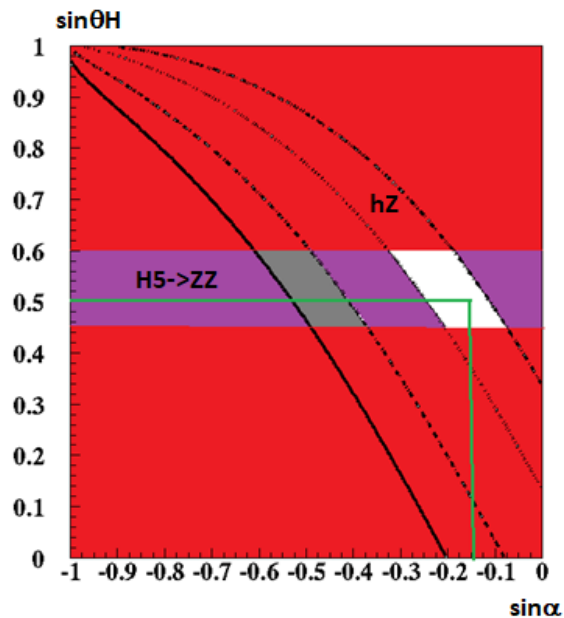

*Figure 15: Constraints from A(400)->hZ (4 curves) and from H5->ZZ (magenta band). Defines two squares, the left-handed one (in grey) being rejected by the h(125)ZZ coupling constraint from figure 11.*

The overall results in a narrow region, shown in figure 15. I have selected :

**$s_H$=0.50 $c_H$=0.87 $s\alpha$=-0.15 $c\alpha$=0.99** with the vacuum expectations **v$\chi$=43 GeV** and **v$\phi$=214 GeV**.

Note that the search for H++(660)->W+W+ by CMS [28] can be translated into the upper bound $\sin\theta_H$<0.32, which seems to contradict the result of figure 15. This interpretation clearly depends on the assumption for mH3. In [29] it was recommended to assume that BR(H5++>W+W+)=100% while, in our case, the dominant decay will occur into H+3W+, hence an increase in the upper bound on $\sin\theta_H$.

## IV.    Isosinglet h' properties

In GM, a heavy isosinglet h' can decay into ZZ/WW, tt, ZA, WH3+ and hh. The fermionic decay width, proportional to $\sin^2\alpha$ is negligible. The hh decay is dominant, as shown in figure 10 in the main text. This result is claimed by [30] with some simplifications. Reference [31], more rigorous, reaches similar conclusions. For h'(600), one predicts BR(hh)=75%, BR(ZZ+WW)=15% and BR(H3W+AZ)=10%.

The experimental search is summarized by the two following plots ([32], [33]).

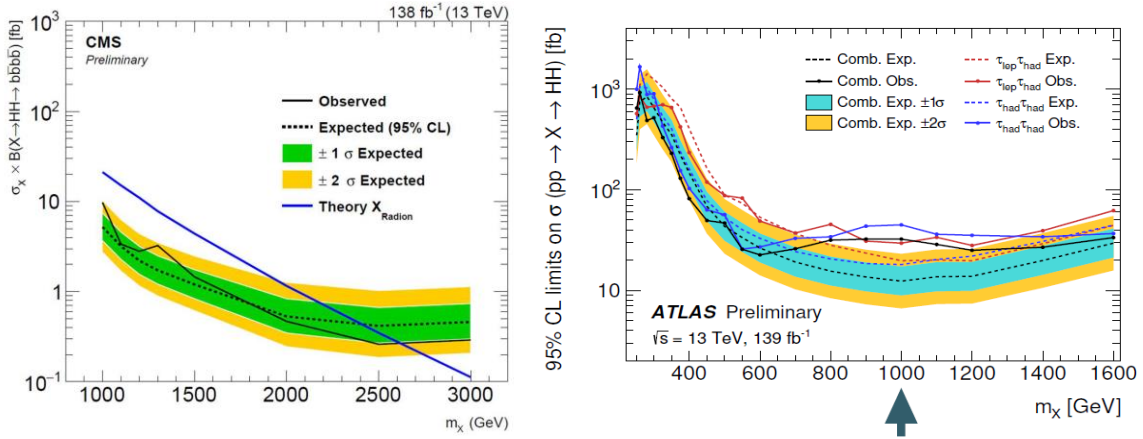

*Figure 16 : Search from ATLAS and CMS for resonances decaying into h(125)h(125).*

ATLAS does not confirm the excess with ττbb in the 4b search, while the 4b search from CMS only starts at 1 TeV, with a suggestive excess in that region. Clearly a mass of 1 TeV goes beyond the expected value on the basis of unitarity bounds (see figure 12).

ATLAS has also observed a ~3.3 s.d. local excess in two photons at 684 GeV, which seems in conformity with the bounds. This excess [34] is concentrated in one bin of 16 GeV, implying a narrow resonance, therefore excluding the wide H5(660) but still allowing for an h'. This would correspond to a cross section σ(pp->h'->γγ) ~0.6 fb. Given the limit set in figure 16, ~25 fb, one easily concludes that it would correspond to a BR(h'-> γγ)~2.4%, that is a Γ(h'->γγ)~400 keV.

This effect is a reminder of the famous/infamous observation of a resonance in two photons at 750 GeV with a 5fb cross section in both experiments, where various interpretations were attempted, including the GM model. Reference [35] concludes that this model is missing by at least an order of magnitude the 5 fb cross section. The present case seems therefore more acceptable given that the observed cross section is ~0.6fb with a smaller mass.

## V.    What about H3(400) ?

In GM one expects mass degeneracy between A(400) and H3+, hence the question: can LHC exclude H3+(400) ?

This particle is seemingly excluded by LHC MSSM searches but this indirect exclusion (figure below) is model dependent and needs to be revised within EGM. The production process gg->H+tb is:

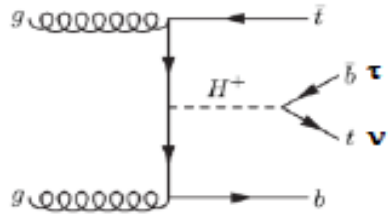

In EGM the xsection goes like $(\zeta bmb)^2+(\zeta tmt)^2$ with $\zeta b\sim 20$   $0.3<\zeta t<0.7$, to be compared to $mb^2t^2\beta+mt^2/t^2\beta$ in MSSM. $0.3<\zeta t<0.7$ corresponds to $1.3<t\beta<2.3$ for MSSM and the direct exclusion from H+->tb can be well below 400 GeV (H->hh needs a reinterpretation in terms of EGM) as can be seen in the figure taken from ATL-PHYS-PUB-2020-006 :

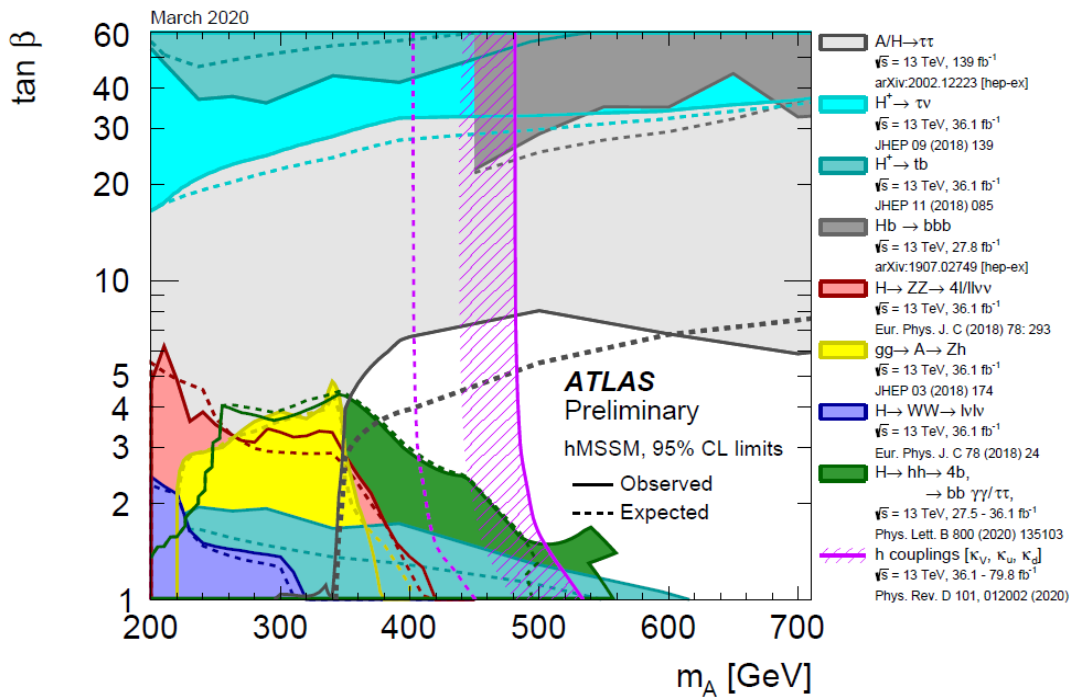